\pdfoutput=1
\documentclass[11pt]{article}

\usepackage[]{ACL2023}

\usepackage{times}
\usepackage{latexsym}

\usepackage[T1]{fontenc}
\usepackage[utf8]{inputenc}

\usepackage{microtype}

\usepackage{inconsolata}

\usepackage{makecell}
\usepackage{enumitem}
\usepackage{graphicx}
\usepackage{mdframed}
\usepackage{amsmath}
\title{Your Instructions Are Not Always Helpful: Assessing the Efficacy of Instruction Fine-tuning for Software Vulnerability Detection}

\author{Imam Nur Bani Yusuf \\
  Singapore Management University \\
  \texttt{imamy.2020@smu.edu.sg} \\
  \And
  Lingxiao Jiang \\
  Singapore Management University \\
  \texttt{lxjiang@smu.edu.sg} \\
  }

\begin{document}
\maketitle
\begin{abstract}
Software, while beneficial, poses potential cybersecurity risks due to inherent vulnerabilities. Detecting these vulnerabilities is crucial, and deep learning has shown promise as an effective tool for this task due to its ability to perform well without extensive feature engineering. However, a challenge in deploying deep learning for vulnerability detection is the limited availability of training data. Recent research highlights the deep learning efficacy in diverse tasks. This success is attributed to instruction fine-tuning, a technique that remains under-explored in the context of vulnerability detection. This paper investigates the capability of models, specifically a recent language model, to generalize beyond the programming languages used in their training data. It also examines the role of natural language instructions in enhancing this generalization. Our study evaluates the model performance on a real-world dataset to predict vulnerable code. We present key insights and lessons learned, contributing to understanding the deep learning application in software vulnerability detection.
\end{abstract}

\section{Introduction}
\label{sec:intro}
Software is integral to various technologies, such as smartphones and e-learning systems. Although software brings many benefits, it presents potential attack surfaces for cyber-attacks due to vulnerabilities. These vulnerabilities, in extreme cases, can lead to significant financial losses. For instance, Revolut experienced a \$20 million loss when attackers exploited weaknesses in its payment system.\footnote{\url{https://thehackernews.com/2023/07/hackers-steal-20-million-by-exploiting.html}} Consequently, the detection of software vulnerabilities is a critical task.

Program analysis techniques, such as static analysis and dynamic analysis, have been used to detect vulnerabilities in code \cite{DBLP:conf/wcre/ZhangDZHYY018, DBLP:conf/issta/ZhangBK19, DBLP:conf/icse/ZhanFCWLLL21}.
However, these techniques suffer from false positives (i.e., detect benign code as vulnerable) or false negatives (i.e., miss many vulnerabilities) \cite{DBLP:journals/tse/ChakrabortyKDR22}. Machine learning has been proposed as an alternative \cite{DBLP:conf/issnip/NarayananCC14, DBLP:conf/mobisys/LiuLJG15, DBLP:conf/icse/MaWGC16}, but it requires extensive feature engineering. Recently, deep learning has emerged as a promising technique due to its good performance without the need for extensive feature engineering \cite{DBLP:conf/ndss/LiZXO0WDZ18, DBLP:conf/sigsoft/FuTLNP22, DBLP:journals/tse/ChakrabortyKDR22}.

The training of deep learning-based vulnerability detection tools faces a significant challenge due to limited data availability. Typically, the instances of vulnerable code are much less frequent than non-vulnerable code. This disparity can lead to a bias towards the majority class, potentially causing the model to overlook the minority class, which in this context is the vulnerable code \cite{DBLP:journals/jbd/JohnsonK19}.

Despite this challenge, recent researches have demonstrated the effectiveness of deep learning in various tasks, including those not encountered during training. This success is largely attributed to instruction fine-tuning, a technique that trains models using task descriptions and corresponding input-label pairs \cite{DBLP:conf/iclr/WeiBZGYLDDL22, DBLP:conf/acl/MishraKBH22, DBLP:conf/acl/HonovichSLS23, DBLP:conf/acl/WangKMLSKH23}. This method leverages task descriptions as a guiding framework for task execution. During fine-tuning, the model learns to match a given task description with specific inputs to generate appropriate outputs. Consequently, the model can infer the correct output based on the provided instruction and input.

Despite its potential, the application of instruction fine-tuning in vulnerability detection remains largely unexplored. This paper aims to investigate how well models can generalize beyond the programming languages present in their training data and examine the impact of natural language instructions on improving generalization in the vulnerability detection task.

We have conducted the study by evaluating a recent language model to predict vulnerable code from the real-world dataset. Our study yields three insights. First, models perform more effectively in scenarios where the language is the same as in the training data (intra-lingual) compared to different languages (cross-lingual), although the difference in performance is not substantial. Second, models that do not use natural language instructions outperform those that do in both intra-lingual and cross-lingual settings. Third, when multiple programming languages are added to the training set without natural language instructions, there's a decline in model performance. Conversely, when natural language instructions are combined with multiple programming languages in the fine-tuning process, the models show better performance.

The contributions of this paper is as follows.
\begin{enumerate}[leftmargin=*, nosep]
    \item We are the first to study how well models can generalize beyond the programming languages present in their training data in the vulnerability detection task.
    \item We are the first to study the impact of natural language instructions on improving generalization in the vulnerability detection task.
\end{enumerate}

The structure of this paper is organized as follows: Section 2 outlines the Research Questions (RQs). Section 3 describes the process of dataset selection. Section 4 elaborates the methodology used for the model selection. Section 5 outlines the experimental settings. Section 6 presents the results. Section 7 discusses the manual case study, lessons learned,  and potential threats to validity. Finally, Section 8 concludes the paper.

\section{Research Questions}
We frame the vulnerability prediction task as a binary classification problem. The classification decision can be formalized as Equation~\ref{eq:problem-formulation}.
\begin{equation}
    v = \arg\max_{v} P(v \mid c)
    \label{eq:problem-formulation}
\end{equation}
The model receives a code snippet $c$ as the input. The objective of the model is to make prediction $v$, where $v$ $\in$ (vulnerable, non-vulnerable). The prediction is based on the conditional probability $P(v|c)$, which represents the likelihood of the code being vulnerable given the code snippet $c$. 

Based on the problem formulation above, this study aims to examine model generalization beyond the languages in the fine-tuning data and the role of natural language instructions in improving generalization performance. We define the following Research Questions (RQs).

\textbf{RQ1: How does the model perform in the intra-lingual and cross-lingual settings?} 
The intra-lingual setting tests the model on the same programming language used during fine-tuning, providing a baseline for performance. The result in the intra-lingual setting will serve a baseline for comparing with the cross-lingual setting. In contrast to the intra-lingual setting, the cross-lingual evaluation tests the model with programming languages different from the fine-tuning language. This question aims to assess the model's ability to generalize across various programming languages.

\textbf{RQ2: How do natural language instructions influence model performance in the intra-lingual and cross-lingual settings?}
This question assesses whether incorporating natural language instructions enhances the model's vulnerability prediction capabilities within the intra-lingual and cross-lingual settings.

\textbf{RQ3: How does the diversity of programming languages in the fine-tuning dataset affect performance in the intra-lingual and cross-lingual settings?}
This question investigates the influence of the programming language diversity in the fine-tuning dataset on the model performance to predict vulnerabilities in the intra-lingual and cross-lingual settings.

\section{Dataset}
\label{sec:dataset-model-selection}

This section describes the dataset selection, statistics, and preprocessing steps.

\subsection{Dataset Selection}
A number of datasets are available in the domain of vulnerability detection, including SARD\footnote{\url{https://samate.nist.gov/SARD/}}, VulDeePecker \cite{DBLP:conf/ndss/LiZXO0WDZ18}, Draper \cite{DBLP:conf/icmla/RussellKHLHOEM18}, D2A \cite{DBLP:conf/icse/ZhengPLBEYLMS21}, Devign \cite{DBLP:conf/nips/ZhouLSD019}, BigVul \cite{DBLP:conf/msr/FanL0N20}, Reveal \cite{DBLP:journals/tse/ChakrabortyKDR22}, DiverseVul \cite{DBLP:conf/raid/0001DACW23}, CrossVul \cite{DBLP:conf/sigsoft/NikitopoulosDLM21}, and CVEFixes \cite{DBLP:conf/promise/BhandariNM21}.

% \begin{table}
% \centering
% \caption{Dataset comparison for vulnerability detection task.}
% \label{tab:dataset-selection}
% \begin{tabular*}{llll}
% \toprule
%      Dataset &    \#Languages &                                        Language &    Label Source \\
% \midrule
%         SARD &  5 &                           C, C++, C\#, Java, PHP &          Synthetic data \\
% VulDeePecker &  2 &                                          C, C++ &           Synthetic data \\
%       Draper &  2 &                                          C, C++ & Annotated by static analyzer \\
%          D2A &  2 &                                          C, C++ & Annotated by static analyzer \\
%       Devign &  2 &                                          C, C++ & Annotated by human \\
%       BigVul &  2 &                                          C, C++ & Curated from security issues \\
%       Reveal &  2 &                                          C, C++ & Curated from security issues \\
%   DiverseVul &  2 &                                          C, C++ & Curated from security issues \\
%     CVEFixes & 28 & C, C++, C\#, JavaScript, Java, Python, Ruby, ... & Curated from security issues \\
%     CrossVul &  5 &            C, JavaScript, PHP, Python, and Ruby & Curated from security issues \\
% \bottomrule
% \end{tabular*}
% \end{table}

VulDeePecker, SARD, BigVul, Draper, D2A, Devign, BigVul, Reveal, and DiverseVul are limited to two programming languages, which are C and C++. Moreover, Draper and D2A are annotated by static analyzers. Hence, Draper and D2A may contain false positive and false negative cases. SARD covers five programming languages. However, SARD is a synthetic dataset and may not reflect real-world vulnerabilities. Similarly, CrossVul and CVEFixes also contain at least five programming languages. However, these datasets reflect real-world vulnerabilities better than SARD because both are curated from real-world security issues. Nevertheless, we choose CVEFixes because it contains a broader range of vulnerability types (i.e., CWE Id) than CrossVul, i.e., 193 vs 167 types.

\subsection{Dataset Preprocessing}
The initial dataset contains 116,057 instances. We exclude instances with more than 990 code tokens to adhere to the maximum context length of 1024 in our experiment, due to the memory constraint. The first exclusion step results in 67,692 instances. Then, we only include dominant languages: C, PHP, and Python. We choose these three languages because they represent different programming use cases and have distinct syntax. Then, we sample randomly 2,750 instances from each language for fine-tuning set and 250 for testing set. We choose this number such that the proportion between the vulnerable and non-vulnerable instances are the same across different programming languages between the fine-tuning and testing sets. In the end, we have 8,250 instances for fine-tuning and 750 instances for testing. Table~\ref{tab:preliminary-dataset} show the statistics for fine-tuning and testing sets. We will use this dataset for the subsequent experiments.

\begin{table}[h]
\centering
\begin{tabular}{l c c}
\hline
\textbf{\makecell[c]{Statistics}} & \textbf{Fine-tuning} & \textbf{Testing} \\
\hline
\#Instances & 8250 & 750 \\
Average \#tokens & 271.80 & 268.12 \\
\#Vulnerable & 4110 & 390 \\
\#Non-vulnerable & 4140 & 360 \\ \hline
\#C & 2750 & 250 \\
\#PHP & 2750 & 250 \\
\#Python & 2750 & 250 \\
\hline
\end{tabular}
\caption{Dataset statistics used in our study. "\#" represents "number of."}
\label{tab:preliminary-dataset}
\end{table}

% C represents lower-level programming languages, PHP is widely used for web development, while Python is for scientific computing. Moreover, they have distinct syntax.

% We randomly sample 3K instances from each language, resulting in 18K instances.  However, this does not guarantee a balanced distribution of vulnerable and non-vulnerable instances. To address this, we further randomly sample 5K instances each of vulnerable and non-vulnerable instances, thus refining the dataset to 9,953 instances. Then, we randomly select 10\% from each language for the testing set and the remaining are used for the fine-tuning set. This ensures an equal representation of each language in both fine-tuning and testing sets. 

\section{Model Selection}
Before conducting the experiments to answer the RQs, we perform a preliminary experiment to select the model with a strong base performance for the vulnerability detection task. We explain the steps as follows.

\subsection{Selecting Model Candidates} 
Models are evaluated against 3 criteria: they must be open-source for custom data fine-tuning, capable of being fine-tuned on a single 24GB memory GPU, and not previously instruction-tuned to ensure performance improvements are due solely to our fine-tuning efforts. From the first filtering, five models are considered: 
CodeLlama (7B) \cite{DBLP:journals/corr/abs-2308-12950}, 
CodeT5+ (770M) \cite{DBLP:journals/corr/abs-2305-07922},
CodeT5 (770M) \cite{DBLP:conf/emnlp/0034WJH21},
Llama (7B) \cite {DBLP:journals/corr/abs-2307-09288},
Mistral (7B) \cite{DBLP:journals/corr/abs-2310-06825}, 
and Yi (6B)\footnote{\url{https://github.com/01-ai/Yi}}, 
CodeLlama, CodeT5+, and CodeT5 are chosen because they are specifically pretrained for code-related tasks. Conversely, Llama 2, Mistral and Yi are chosen as representatives of models pre-trained on general corpora. The selection excludes the larger variants of CodeT5+ due to the lack of optimization with Flash Attention 2 \cite{DBLP:journals/corr/abs-2307-08691}, which restricts fine-tuning on our GPU.

\subsection{Preliminary Experimental Setting}
We fine-tune and evaluate the model using all languages in our dataset, which includes C, PHP, and Python, to identify a model with a strong base performance. We leverage accuracy as the evaluation metric. We compute accuracy by calculating the ratio of correctly predicted instances to the total number of predictions made. This metric is chosen because it reflects the model's ability to accurately predict both vulnerable and non-vulnerable instances. Ideally, the base model should demonstrates high performance in both classes. Note that the testing set has been balanced to contain an equal proportion of vulnerable and non-vulnerable instances, ensuring that accuracy reflects an unbiased assessment of the model's predictive capabilities.

\subsection{Preliminary Experiment Result}

\begin{table}[h]
\centering
\begin{tabular}{l c}
\hline
\textbf{\makecell[c]{Model}} & \textbf{Accuracy} \\
\hline
CodeLlama & \textbf{0.535} \\
CodeT5+ & 0.520 \\
CodeT5 & 0.480 \\
Llama 2 & 0.511 \\
Mistral & 0.531 \\
Yi & 0.480 \\
\hline
\end{tabular}
\caption{The preliminary experiment result. CodeLlama performs the best among the others.}
\label{tab:preliminary-results}
\end{table}

Table~\ref{tab:preliminary-results} reveals the results of the preliminary experiments. CodeLlama performs the best among the other models with accuracy 0.535. Therefore, we select CodeLlama as the base model to answer our Research Questions (RQs) in the subsequent experiments.

% \textbf{Dataset Preprocessing.}
% The initial dataset contains 116,057 instances. We exclude instances with more than 990 code tokens to adhere to the maximum context length of 1024, resulting in 67,692 instances. Then, we only include dominant languages: C, C++, Java, JavaScript, PHP, and Python. We randomly sample 3K instances from each language, resulting in 18K instances.  However, this does not guarantee a balanced distribution of vulnerable and non-vulnerable instances. To address this, we further randomly sample 5K instances each of vulnerable and non-vulnerable instances, thus refining the dataset to 9,953 instances. Then, we randomly select 10\% from each language for the testing set and the remaining are used for the fine-tuning set. This ensures an equal representation of each language in both fine-tuning and testing sets. In the end, we have 8,999 instances for fine-tuning and 954 instances for testing. Table~\ref{tab:preliminary-dataset} show the statistics for fine-tuning and testing sets.

\section{Study Experimental Setting}
This section discusses the pipeline in the fine-tuning and inference stages. Then, this section elaborates the implementation details, including the model implementation, training setting, and how to generate the natural language instructions.

\subsection{Fine-tuning}
% The input of the fine-tuning stage is a pair of code $c$ and the corresponding label $v$. First, the tokenizer tokenizes the code $c$ into a sequence of token ids $t$. Then, the embedding layer converts each token id $t_i \in t$ into a sequence of vectors $t'$. The transformer blocks (either encoder or decoder, depends on the model architecture) receives these vectors and outputs the probability of the whole tokens in the vocabulary. Then, the model predicts the next token by taking the argmax. The loss function receives the predicted and ground truth next tokens as the input and compute the loss to optimize the model. We leverage standard loss function for the next token prediction task, which is Cross Entropy~\cite{cox1958regression}.

% Let $c$ represent the input and $v$ its corresponding label in the fine-tuning process. A template function $Q$ is applied to $c$ and $l$, transforming it into $w=Q(c, l)$. The function $Q$ formats $c$ and $l$ into a specific template, depending whether the experiment leverages natural language instructions or not. The template is shown in Figure~\ref{fig:template}. 

Let \( c \) represent the input and \( v \) its corresponding label in the fine-tuning process. A transformation is applied using a template function \( Q \), which takes \( c \) and a label \( l \) as inputs and produces an output \( w \). This output is denoted as \( s = Q(c, l) \). The function \( Q \) is responsible for formatting \( c \) and \( l \) into a specific template. The choice of template depends on whether the experiment incorporates natural language instructions. The details of this template are depicted in Figure~\ref{fig:template}.

\begin{figure}[h]
    \centering
    \begin{mdframed}[linecolor=black, middlelinewidth=2pt]
        \underline{with instruction} \newline
        <s> [INST] \{insert instruction\} [/INST] \newline
        \#\#\# Code: \{insert code\} \newline
        \#\#\# Answer: \{insert label\} </s> \newline

        \underline{without instruction} \newline
        \#\#\# Code: \{insert code\} \newline
        \#\#\# Answer: \{insert label\} </s>
    \end{mdframed}
    \caption{The distinct formats for the template function $Q(c, l)$, showing variations with and without the inclusion of natural language instructions.}
    \label{fig:template}
\end{figure}

Given a sequence \( s \), a function \( F \) generates a set of input-label pairs, \( F(s) = \{ ((s_1, \dots, s_j), s_{j+1}) \mid 1 \leq j < n \} \), where each pair consists of a subsequence of \( s \) and the following word as the label. These pairs are then processed in a batch where a tokenizer function \( \tau \) converts each item in the batch into a sequence of token ids \( t = \tau(w) \). An embedding function \( \epsilon \) maps each token id \( t_i \in t \) to a sequence of vectors \( t' = \epsilon(t) \). The transformer architecture (either encoder or decoder, depending on the model), represented as a function \( \phi \), processes \( t' \) to produce a probability distribution over the vocabulary for each token, denoted as \( P = \phi(t') \). The prediction of the next token \( \hat{t}_{i+1} \) is determined by \( \hat{t}_{i+1} = \operatorname{argmax}(P) \). A loss function \( L \), i.e., Cross Entropy~\cite{cox1958regression}, is used to compute the loss. It takes as input the predicted token \( \hat{t}_{i+1} \) and the ground truth next token \( t_{i+1} \), yielding the loss value \( L(\hat{t}_{i+1}, t_{i+1}) \) to optimize the model.

% A tokenizer function $\tau$ converts $w$ into a sequence of token ids $t=\tau(w')$. An embedding function  $\epsilon$ maps each token id $t_i \in t$ to a sequence of vectors $t'=\epsilon(t)$. The transformer architecture (either encoder or decoder, depending on the model), represented as a function $\phi$, processes $t'$ to produce a probability distribution over the vocabulary for each token, denoted as $P = \phi(t')$. The prediction of the next token $\hat{t}_{i+1}$ is determined by $\hat{t}_{i+1} =\operatorname{argmax}(P)$. A loss function $L$, i.e., Cross Entropy~\cite{cox1958regression}, is used to compute the loss. It takes as input the predicted token $\hat{t}_{i+1}$ and the ground truth next token $t_{i+1}$, yielding the loss value  $L(\hat{t}_{i+1}, t_{i+1})$ to optimize the model.

\subsection{Inference}
Given a code snippet \( c \), the transformation function \( Q \) formats the code snippet \( c \) into a specific format, resulting in \( w = Q(c) \). This process differs from the training stage, as the label \( v \) is not included in the input to \( Q \). Next, the tokenizer function processes \( w \) to produce a sequence of token ids \( t = \tau(w) \). The embedding layer converts these token ids into vectors \( t' = \epsilon(t) \). The transformer block takes these vectors and computes a probability distribution \( P = \phi(t') \) over the entire vocabulary. Notably, the token with the highest probability, \( \operatorname{argmax}(P) \), may not match the target labels 'True' or 'False'. To address this, we only extract the probabilities of the 'True' and 'False' tokens, i.e., \( P_{\text{True}} \) and \( P_{\text{False}} \). The final output is the token with the greater probability, determined by \( \text{max}(P_{\text{True}}, P_{\text{False}}) \).

\subsection{Evaluation Metrics}
We choose the F1 score as the evaluation metrics because it has been widely used to evaluate vulnerability detection tools~\cite{DBLP:conf/msr/FuT22, DBLP:conf/icse/SteenhoekRJL23, DBLP:journals/corr/abs-2212-08108}. The F1 score considers both the precision and the recall of the test to compute the score. Precision is the number of correct positive results  divided by the number of all positive results, and recall is the number of correct positive results divided by the number of positives that should have been returned. Here, positive results refers to vulnerable code. The F1 score is the harmonic mean of precision and recall, giving both measures equal weight.

\subsection{Implementation Details}
\subsubsection{Models and Training Setting} 
For all models, we use the official implementation from HuggingFace. The link to the model can be found in the Appendix.

We fine-tune the model using the LoRA technique~\cite{DBLP:conf/iclr/HuSWALWWC22} with \( r = 16 \) and \( \alpha = 32 \), where \( r \) is chosen as the largest value that does not trigger an out-of-memory error and \( \alpha \) is double the value of \( r \), as recommended.  To accommodate the model within our hardware limitations, we implement several optimizations: loading the model with 4-bit precision, limiting the maximum context length to 1024 tokens, utilizing Flash Attention 2, and applying gradient checkpointing. The learning rate is set to \( 3 \times 10^{-4} \) and the batch size is chosen as the largest multiple of 2 that fits our machine, i.e., \( 8 \). We employ a cosine learning rate scheduler with warmup steps of 100 to mitigate unstable learning that may result from arbitrary initialization of optimizer states. We conduct our fine-tuning process on a machine with Linux 22.04.2 and an RTX 3090 GPU with 24GB of memory. We perform all experiments three times with three different random seeds to avoid a lucky trial.

\subsubsection{Generating and Choosing Instructions}
Initially, we create a set of instructions manually. These initial instructions are then refined using GPT-4, yielding 15 refined versions. From these, we select 5 instructions for our final instruction pool. This selection is based on evaluating the quality and diversity through human judgment. The rationale for limiting to 5 instructions is to prevent an excessive number that might impede the model's learning ability, potentially treating the instructions as noise. During the experiment using natural language instructions (RQ3), the template function $Q$ chooses one instruction from the instruction pool using random sampling.
\section{Experimental Results}

% This section describes the experiment result for each RQ.

% \begin{figure*}[t]
%     \centering
%     \includegraphics[width=\textwidth]{intra-cross-no-instruct.png}
%     \caption{The results for intra-lingual and cross-lingual settings.}
%     \label{fig:intra-cross-results}
% \end{figure*}

\begin{figure*}
    \centering
    \includegraphics[width=\textwidth]{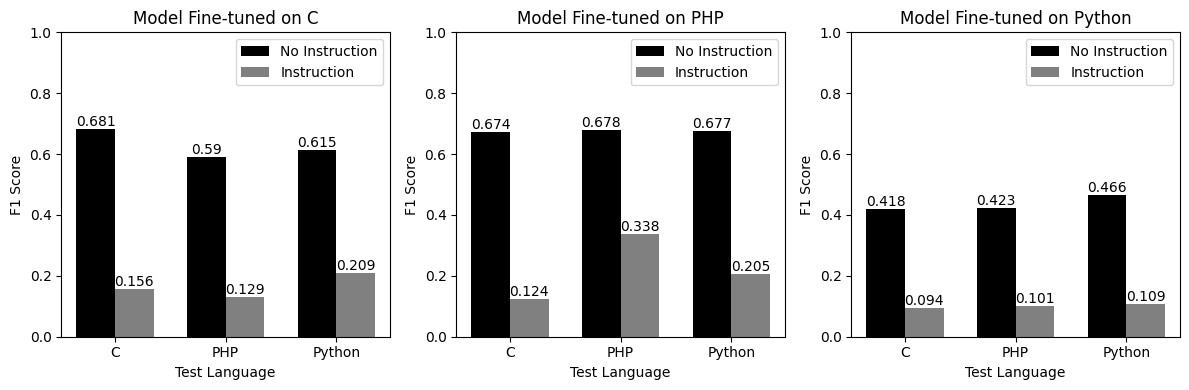}
    \caption{The F1 scores for models fine-tuned on C, PHP, and Python, across different test languages. Two sets of bars are displayed for each test language, representing model performance with (grey) and without (black) natural language instructions.}
    \label{fig:intra-cross-results}
\end{figure*}

Figure~\ref{fig:intra-cross-results} illustrates the model's performance in the intra-lingual and cross-lingual settings. In the intra-lingual context, the C$\rightarrow$C model (i.e., the model fine-tuned and tested on C) exhibits the highest performance, closely followed by the (PHP$\rightarrow$PHP) and (Python$\rightarrow$Python) models. When comparing intra-lingual to cross-lingual performance for each programming language, models generally show better results in the intra-lingual setting. For example, the C$\rightarrow$C model outperforms the C$\rightarrow$PHP and C$\rightarrow$Python models. This pattern is consistent when the model is tested using PHP and Python languages. A higher performance in the intra-lingual setting is expected, given the syntactical similarities between the fine-tuning and testing datasets. 

However, an interesting observation is the relatively small performance gap between intra-lingual and cross-lingual settings. For example, the PHP$\rightarrow$PHP model exhibits similar performance to the PHP$\rightarrow$C model and the PHP$\rightarrow$Python model. A similar trend is also observed in the C$\rightarrow$C model against the C$\rightarrow$PHP and C$\rightarrow$Python models and Python$\rightarrow$Python against the Python$\rightarrow$C and Python$\rightarrow$PHP models.

\begin{mdframed}[linecolor=black, middlelinewidth=2pt]
\textbf{Answer to RQ1:} The models perform better in the intra-lingual than cross-lingual settings, as shown by Figure~\ref{fig:intra-cross-results}. However, while cross-lingual performance is generally lower, the gap between intra-lingual and cross-lingual results is notably small.
\end{mdframed}

Figure~\ref{fig:intra-cross-results} also presents the evaluation results for intra-lingual and cross-lingual settings, comparing scenarios with and without natural language instructions. A significant decline in the model's performance is evident when inputs include natural language instructions. Such a case is indicated by the lower F1 scores of grey bars compared to black bars in Figure~\ref{fig:intra-cross-results}. For instance, the model fine-tuned and tested on C without instructions C$\rightarrow$C yields better performance than the model fine-tuned and tested on C with instructions (C\textsubscript{instruct}$\rightarrow$C\textsubscript{instruct}). The same result is also observed in the Python$\rightarrow$Python model against the Python\textsubscript{instruct}$\rightarrow$Python\textsubscript{instruct} model and the PHP$\rightarrow$PHP model against the PHP\textsubscript{instruct}$\rightarrow$PHP\textsubscript{instruct} model. The cross-lingual setting also exhibits the same trend.  This decrease in performance might be due to the natural language instructions inadvertently diverting the model's attention from important syntactical and semantic features of the code, which are essential to predict the existence of vulnerabilities in code.

\begin{mdframed}[linecolor=black, middlelinewidth=5pt]
\textbf{Answer to RQ2:} The models without natural language instructions outperform those with instructions across intra-lingual and cross-lingual settings. This is exemplified by higher F1 scores in Figure~\ref{fig:intra-cross-results} in scenarios such as C$\rightarrow$C compared to C\textsubscript{instruct}$\rightarrow$C\textsubscript{instruct}, and similarly for Python and PHP models. 
% The consistent trend across different languages suggests that natural language instructions may interfere with the model's ability to focus on crucial code characteristics for vulnerability prediction. This insight calls for a reconsideration of the role of natural language instructions in the fine-tuning process of such models.
\end{mdframed}

\begin{figure*}[t]
    \centering
    \includegraphics[width=\textwidth]{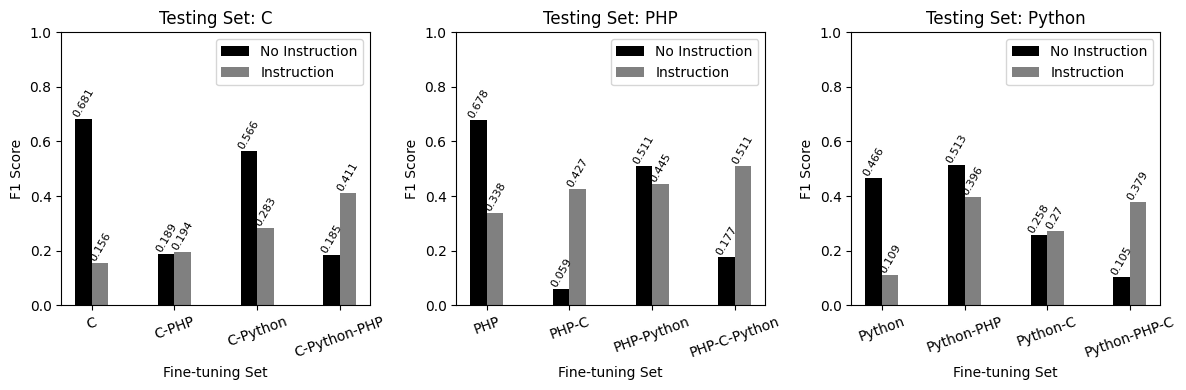}
    \caption{Comparative Performance of Models in Different Testing Sets. These grouped bar charts illustrate the F1 scores for models fine-tuned on various language combinations (C, C-PHP, C-Python, C-Python-PHP) and tested on C, PHP, and Python. Each chart contrasts the model's performance with (grey bars) and without (black bars) natural language instructions.}
    \label{fig:dataset-diversity}
\end{figure*}

Figure~\ref{fig:dataset-diversity} demonstrates the impact of language diversity in the fine-tuning set on the model's performance. For the model tested on the C language (the leftmost chart of Figure~\ref{fig:dataset-diversity}), incorporating additional languages when fine-tuning the model without natural language instructions leads to a decrease in the model's performance. This finding aligns with our expectations. However, the results show that adding relevant languages when fine-tuning the model without natural language instructions does not enhance performance but yields a significant decline. Such a case is indicated by the higher black bars in Figure~\ref{fig:dataset-diversity} when the model is fine-tuned using a single language rather than multiple languages. For example, in the middle chart of Figure~\ref{fig:dataset-diversity}, the C-PHP$\rightarrow$PHP model exhibits poorer performance than the C$\rightarrow$PHP model. The same trend is also observed in the C-Python$\rightarrow$Python model against the C$\rightarrow$Python model. Similarly, including all languages (C, PHP, and Python) when fine-tuning the model without natural language instructions also does not help the model perform better than only fine-tuning the model using a single language.

Conversely, a different pattern emerges when fine-tuning using natural language instructions with more programming languages. Adding more programming languages can improve the model's performance across all test sets. This improvement is illustrated by the higher grey bars in Figure~\ref{fig:dataset-diversity} when the model is fine-tuned using multiple languages than when only using a single language. For instance, the C-Python-PHP$\rightarrow$C, C-Python$\rightarrow$C, and C-PHP$\rightarrow$C models demonstrate better performance than the C$\rightarrow$C model as indicated by the leftmost chart of Figure~\ref{fig:dataset-diversity}. The PHP and Python languages also exhibit the same trend.

\begin{mdframed}[linecolor=black, middlelinewidth=2pt]
\textbf{Answer to RQ3:}  Adding multiple programming languages to the fine-tuning set without natural language instructions reduces model performance. This is shown by higher black bars in the charts in Figure~\ref{fig:dataset-diversity} for models fine-tuned on multiple languages compared to those on a single language.  Conversely, incorporating natural language instructions and multiple languages in the fine-tuning process improves the model's performance, as indicated by higher grey bars in the charts.
\end{mdframed}

\section{Discussion}
This section discusses the manual case study, lessons learned, and threats to validity of this study.

\subsection{Can Language-specific Instructions Improve the Model's Performance?}
One hypothesis for the model's poor performance when the model is fine-tuned using natural language instructions is that the instruction pool lacks language-specific cues, i.e., the same instructions are used across all languages. To verify this, we conducted a check by modifying the experimental setup from RQ3. We redo the experiments by explicitly stating the programming language of the code snippet in the instruction, as illustrated in the updated template shown in Figure~\ref{fig:template-new}. We conduct the new experiments three times using three different random seeds.

\begin{figure}[h]
\centering
\begin{mdframed}[linecolor=black, middlelinewidth=2pt]
The following code is written in \{insert language\} \newline
<s> [INST] \{insert instruction\} [/INST] \newline
\#\#\# Code: \{insert code\} \newline
\#\#\# Answer: \{insert label\} </s> \newline
\end{mdframed}
\caption{The updated template that incorporates language information in the instruction.}
\label{fig:template-new}
\end{figure}

\begin{table}[h]
\centering
\begin{tabular}{lccc}
\hline
\textbf{\makecell[c]{Fine-tuning set}} & \textbf{F1\_new} & \textbf{F1\_ori} \\ \hline
 C                  & 0.054            & 0.156             \\ 
 C-PHP              & 0.081            & 0.194             \\ 
 C-Python           & 0.057            & 0.283             \\ 
 C-Python-PHP       & 0.045            & 0.411             \\ \hline
\end{tabular}
\caption{The comparison results between the updated and old templates when tested on C languages. F1\_ori is the F1 score using the instruction in Figure~\ref{fig:template}, while F1\_new is the F1 score using the instruction in Figure~\ref{fig:template-new}.}
\label{tab:result_new_instruction}
\end{table}

Table~\ref{tab:result_new_instruction} presents a comparison between the new results obtained using the updated template and the previous results from Figure~\ref{fig:dataset-diversity}. These results demonstrate a performance decline when employing the updated template. Specifically, the F1 scores for models tested on the C language with instructions from Figure~\ref{fig:template-new} are lower than those achieved with the original instruction format in Figure~\ref{fig:template}. A similar trend is observed in tests involving PHP and Python languages. This pattern suggests that adding more instructions might detrimentally affect model performance. Furthermore, the significant performance drop highlighted in Table~\ref{tab:result_new_instruction} indicates a notable sensitivity of the model to the instructions.

\subsection{Is the Model Fine-tuned on Language X Able to Predict a Similar Vulnerability in Language Y?}
We investigate whether models fine-tuned in one programming language can predict vulnerabilities in another by comparing intra-lingual and cross-lingual settings within the same testing language. Specifically, we chose the C$\rightarrow$C model without instructions for intra-lingual comparison due to its superior performance as shown in Figure~\ref{fig:intra-cross-results}. For cross-lingual comparison, we analyze models trained on a single language but tested on C, namely the PHP$\rightarrow$C and Python$\rightarrow$C models. Both models are also fine-tuned without instructions. We compare the top-3 correct predictions across these models, which are consistent across all: CWE-125 (out of bounds read), CWE-119 (buffer overflow), and CWE-20 (improper input validation), as detailed in Table~\ref{tab:correct-preds-cwe}.

\begin{table}[h]
\centering
\begin{tabular}{lccc}
\hline
\makecell[c]{CWE-ID} & C$\rightarrow$C & PHP$\rightarrow$C & Python$\rightarrow$C \\ \hline
CWE-125         & 36              & 36                & 19                   \\ 
CWE-119         & 23              & 23                & 7                    \\ 
CWE-20          & 20              & 20                & 6                    \\ \hline
\end{tabular}
\caption{Number of correct instances for the top-3 predicted vulnerabilities for the C$\rightarrow$C, PHP$\rightarrow$C, and Python$\rightarrow$C models.}
\label{tab:correct-preds-cwe}
\end{table}

Further analysis is conducted by examining the fine-tuning data for occurrences of the CWEs listed in Table~\ref{tab:correct-preds-cwe}, as summarized in Table~\ref{tab:cwe-in-finetuning}.

\begin{table}[h]
\centering
\begin{tabular}{lccc}
\hline
\makecell[c]{CWE-ID} & {C} & {PHP} & {Python} \\ \hline
CWE-125         & 353        & 1            & 68              \\ 
CWE-119         & 274        & 0            & 26              \\ 
CWE-20          & 230        & 136          & 153             \\ \hline
\end{tabular}
\caption{Number of instances for each CWE-ID in the fine-tuning dataset.}
\label{tab:cwe-in-finetuning}
\end{table}

The data reveals that models can recognize similar vulnerabilities even when fine-tuned in different languages. For example, the Python$\rightarrow$C model, initially fine-tuned with 68 instances of CWE-125 in Python, successfully detects 19 instances in C, as shown in Table~\ref{tab:correct-preds-cwe}. Notably, the PHP$\rightarrow$C model, despite being fine-tuned with only one instance of CWE-125, matches the C$\rightarrow$C model's performance in correctly detecting 36 vulnerability instances in C language. Moreover, even with no instances of CWE-119 in its fine-tuning data, the PHP$\rightarrow$C model correctly predicts 36 instances when evaluated in C language. We conduct a manual check and find that CWE-119 and CWE-125 are closely related, i.e., buffer overflow can potentially cause out-of-bound access. This could suggest a correlation between CWE-119 and CWE-125, where learning to detect one may improve the detection of the other.

\subsection{Lessons Learned}
This study yields several insightful lessons, which are outlined as follows:

\textbf{Instruction-based fine-tuning may not always enhance performance.} Our findings suggest that fine-tuning language models using natural language instructions does not invariably lead to improved performance. This observation contrasts with prior studies~\cite{DBLP:conf/iclr/WeiBZGYLDDL22, DBLP:conf/acl/HonovichSLS23}, which highlighted the benefits of instruction-based fine-tuning. Consequently, we encourage the research community to further investigate this area to gain a deeper understanding of instruction-based fine-tuning.

\textbf{Single-language fine-tuning outperforms multi-language fine-tuning.} Our findings reveal that training language models with data from multiple languages can significantly diminish their performance. Therefore, it appears more effective to fine-tune using data from a single language. This can be attributed to several factors, such as syntax and grammatical complexity, limited model capacity, and cross-lingual interference.

\textbf{Predicting similar vulnerabilities across languages.} Our results indicate the feasibility of accurately predicting similar vulnerabilities in different programming languages. This approach could be particularly advantageous when dealing with low-resource languages. An alternative strategy might involve identifying languages with a diverse range of vulnerabilities and using these as the basis for fine-tuning, rather than relying solely on data from low-resource languages.

\subsection{Threats to Validity}
\textbf{Internal Validity:} The choice of the model can significantly influence our study. To minimize such threats, we systematically select the model through initial filtering and preliminary experiments. Additionally, we employ the official implementations available on HuggingFace to ensure consistency. We also conduct the experiments three times, which helps to avoid bias that might result in a particular trial favoring specific settings.

\textbf{External Validity:} The selection of our dataset can impact the study. To minimize this threat, we choose the dataset systematically. Furthermore, the class imbalance between vulnerable and non-vulnerable code is a potential issue. We address this by carefully preprocessing the dataset to balance the number of vulnerable and non-vulnerable code snippets across different programming languages. We also ensure equal representation of each programming language in the dataset. However, our study is limited to three languages: C, PHP, and Python, considering the time and resource limitations. However, we consider their syntax diversity and the programming use cases to make sure that the chosen languages are representative. Moreover, the threat of randomness is mitigated by setting a fixed random seed for each experiment trial.

\textbf{Construct Validity:} The choice of evaluation metrics can potentially threaten the validity of our study, as inappropriate metrics may not accurately capture the intended outcomes. To minimize this threat, we conduct a manual case study alongside the automated metrics in Section 7.

\section{Conclusion and Future Work}
This study scrutinizes the model generalization beyond the languages in the fine-tuning data and the role of natural language instructions in improving generalization performance in the vulnerability detection task. We have conducted experiments using recent models to detect real-world vulnerabilities. Our study yields three insights. First, models perform more effectively in scenarios where the language is the same as in the training data (intra-lingual) compared to different languages (cross-lingual), although the difference in performance is not substantial. Second, models that do not use natural language instructions outperform those that do in both intra-lingual and cross-lingual settings. Third, when multiple programming languages are added to the training set without natural language instructions, there's a decline in model performance. Conversely, when natural language instructions are combined with multiple programming languages in the fine-tuning process, the models show better performance.

\textbf{Future Work.} Future research can further investigate the generalization capabilities of models in identifying different types of vulnerabilities. Our initial case study indicates that the model can generalize from one vulnerability type, such as a buffer overflow, to a closely related one, like out of bounds read. However, comprehensive studies are required to confirm its applicability to a broader range of vulnerabilities. Additionally, this work has primarily focused on straightforward fine-tuning instructions. Exploring advanced techniques like Chain-of-Thought~\cite{DBLP:conf/nips/Wei0SBIXCLZ22}, instruction with reasoning~\cite{DBLP:journals/corr/abs-2311-11045}, or instruction with demonstration~\cite{DBLP:conf/icml/ShaoGSHDC23}, which have shown promise in other tasks, could offer valuable insights in the context of vulnerability detection. Finally, expanding experiments to include a diverse array of models and programming languages could provide a more comprehensive understanding of the model's effectiveness across different scenarios.

\bibliography{anthology,custom}
\bibliographystyle{acl_natbib}

% \appendix

% \section{Example Appendix}
% \label{sec:appendix}

% This is a section in the appendix.

\end{document}